# Electrowetting on liquid-infused film (EWOLF): Complete reversibility and controlled droplet oscillation suppression for fast optical imaging

Chonglei Hao[1], Yahua Liu[1], Xuemei Chen[1], Yuncheng He[2], Qiusheng Li[2], K.Y. Li[1], Zuankai Wang[1,*]

[1]Department of Mechanical and Biomedical Engineering,
City University of Hong Kong, Hong Kong, China

[2]Department of Civil and Architectural Engineering,
City University of Hong Kong, Hong Kong, China

*Corresponding author. E-mail: zuanwang@cityu.edu.hk

**Abstract**

Electrowetting on dielectric (EWOD) has emerged as a powerful tool to electrically manipulate tiny individual droplets in a controlled manner. Despite tremendous progress over the past two decades, current EWOD operating in ambient conditions has limited functionalities posing challenges for its applications, including electronic display, energy generation, and microfluidic systems. Here, we demonstrate a new paradigm of electrowetting on liquid-infused film (EWOLF) that allows for complete reversibility and tunable transient response simultaneously. We determine that these functionalities in EWOLF are attributed to its novel configuration, which allows for the formation of viscous liquid-liquid interfaces as well as additional wetting ridges, thereby suppressing the contact line pinning and severe droplet oscillation encountered in the conventional EWOD. Finally, by harnessing these functionalities demonstrated in EWOLF, we also explore its application as liquid lens for fast optical focusing.

Electrowetting on dielectric (EWOD), owing to its ability to electrically manipulate tiny individual droplets without involving movable mechanical parts, has received much attention in the past two decades[1-14]. Different from the configuration of classical mercury/electrolyte electrocapillary system[15,16], EWOD devices typically rely on the sandwiching of a solid dielectric layer between a metallic electrode and aqueous electrolyte[17-19]. When an electric field is applied across the dielectric layer, liquid-solid surface tension is tailored and contact angle (CA) of electrolyte droplet is decreased. The presence of the dielectric layer not only eliminates unwanted electrolysis, but also makes it possible to achieve a large switching angle. Owing to its low energy consumption, favorable scaling of the actuation force at the microscale as well as amenable integration with other systems, EWOD has also found promising applications in electronic display[20-22], energy generation[23], microelectromechanical (MEMS)[24,25] and microfluidic systems[26-29].

Despite tremendous promise, the use of solid dielectric layer between the aqueous droplet and underlying electrode is associated with inevitable physical and chemical heterogeneities[30-32], leading to limited functionalities. For example, owing to the large CA hysteresis, contact line pinning[30-33] as well as CA saturation[34-36] at high voltage, it remains a challenge to achieve reversible electrowetting with a large degree of switchability in ambient conditions. Moreover, activating droplet in EWOD is vulnerable to pronounced oscillation in response to an abrupt external stimulus, resulting in elongated time for the droplet to reach its equilibrium state[37]. Such undesired transient response instability, together with the limited reversibility is not compatible with many desired functions, such as fast real-time focusing systems[38-41]. Thus, there is a pressing need to develop novel strategies to address the challenges encountered in the EWOD.

In this work, we report a new electrowetting approach that features a liquid-infused film as the dielectric layer to impart complete reversibility and droplet oscillation suppression during the transient response. The liquid-infused film is achieved by locking a liquid lubricant in a porous membrane through the delicate control of wetting properties of the liquid and solid phases. Taking advantage of the negligible contact line pinning at the liquid-liquid interface[42-47], the droplet response in EWOLF can be electrically addressed with full switchability and reversibility compared to the conventional EWOD on superhydrophobic surfaces[2,31,48]. Moreover, we show that the infiltration of liquid lubricant phase in the porous membrane also efficiently enhances the viscous energy dissipation, suppressing the droplet oscillation and leading to fast response without sacrificing the desired electrowetting reversibility. Meanwhile, we find that the damping effect associated with the EWOLF can be tailored by manipulating the viscosity and thickness of liquid lubricant. We also demonstrate the feasibility of developing adaptive liquid lens for fast optical focusing using the as-proposed EWOLF.

**Results**

**Sample preparation.** The basic component of the EWOLF is a thin liquid-infused dielectric film, which is constructed by infiltrating lubricating oil in a porous polytetrafluoroethylene (PTFE) membrane. Recently, the liquid-infused film has been elegantly demonstrated for a number of applications including enhanced dropwise condensation, anti-icing and anti-bacterials[43,49-51].

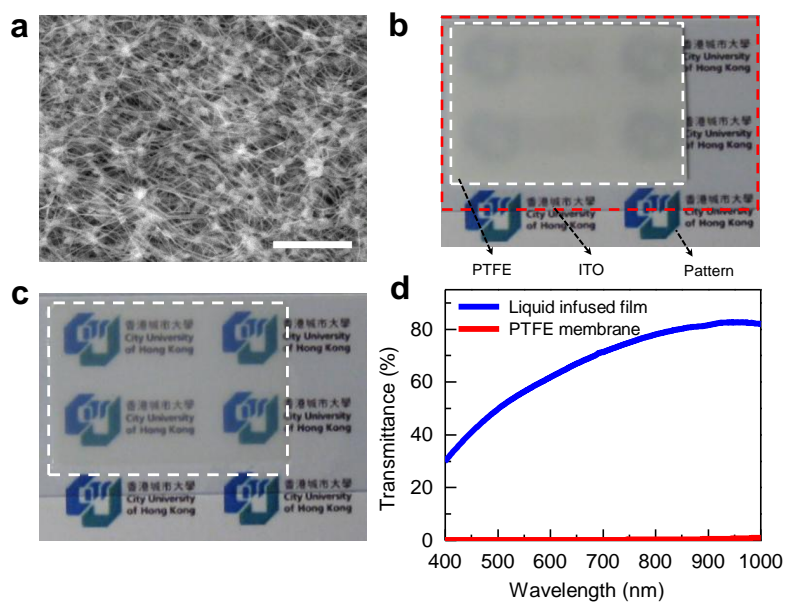

**Figure 1 | SEM image of the PTFE membrane, optical images and transmittance measurement of the PTFE membrane and liquid-infused film.** (a) SEM image of the PTFE membrane consisting of nano-fiber networks. The scale bar is 4 μm. (b) Photograph showing a PTFE membrane (white dashed-line rectangle) attached to a transparent ITO glass substrate (red dashed-line rectangle). Underneath the PTFE/ITO is a paper with printed patterns (logo arrays, in blue and green). Without the infiltration of oil lubricant, the PTFE membrane is opaque. (c) Optical image of the liquid-infused film (circled with white dashed-lines) attached to an ITO substrate. The pattern underlying the film is clearly visualized, revealing

the transparency of the liquid-infused film. **(d)** Optical transmittance measurement showing the enhanced optical transmittance of the liquid-infused film (blue solid line) relative to that of the PTFE membrane (red solid line). The enhanced optical transmittance in the liquid-infused film is owing to the reduced light scattering at the air/liquid/solid interfaces rendered by the replacement of air pockets by the oil.

A detailed description of the preparation of the liquid-infused film is shown in the Methods section. Fig. 1a shows a characteristic scanning electron microscope (SEM) image of PTFE membrane. The average pore size, thickness, and solid fraction of the PTFE membrane are ~ 200 nm, ~ 20 μm and ~ 0.16, respectively. The optical image of a PTFE membrane attached to a transparent indium tin oxide (ITO) (~ 180 nm) coated glass substrate is shown in Fig. 1b. Underlying underneath the PTFE/ITO/glass is a paper with printed logo arrays (in blue and green). Obviously, these logo arrays are hazy, indicating the PTFE membrane's inability to transmit light. Interestingly, after the infiltration of liquid lubricant into the PTFE membrane, the logo arrays are clearly visualized (Fig. 1c). According to the optical transmittance measurement using Perkin Elmer Lanbda 35 UV-VIS Spectrometer as shown in Fig. 1d, the liquid-infused film exhibits excellent optical transmittance in the visible light wavelength range whereas the optical transmittance of the PTFE membrane is nearly zero. The enhanced optical transmittance in the liquid-infused film is due to reduced light scattering at the air/liquid/solid interfaces rendered by the replacement of air pockets by the liquid lubricant[43,45]. The thickness of the liquid-infused film is ~ 50 μm, which can be tailored by varying the amount of lubricant infiltrated into the PTFE membrane. The water apparent CA on the PTFE membrane is ~ 152 °, indicating its superhydrophobic wetting property. In contrast, the apparent CA of water on the liquid-infused film is ~ 103 °.

**Complete reversibility of EWOLF.** The hybrid liquid-infused dielectric layer is sandwiched between a conductive ITO substrate and water droplet. The droplet is connected to an extremely slim Tungsten tip with diameter of only ~ 25 μm, which serves to establish an electric potential difference with the underlying ITO electrode.

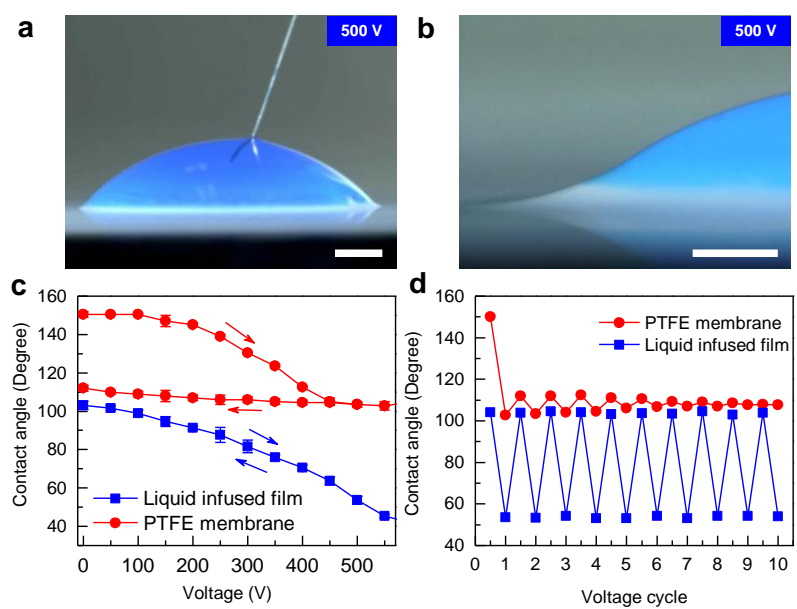

**Figure 2 | Wetting properties of liquid-infused film and electrowetting response. (a)** Optical image of a stained droplet in EWOLF subject to an actuation voltage of 500 V, which displays an apparent CA of ~53 °. The scale bar is 400 μm. **(b)** The formation of wetting ridge as a result of oil motion at the liquid-liquid interface. The scale bar is 200 μm. **(c)** Characterization

of the variations of apparent CA in EWOD and EWOLF. For the liquid-infused film, there is a perfect fidelity between the apparent CA in the spreading (the applied potential is increased) and receding (the applied potential is decreased) period, indicating a full reversibility. In contrast, there exists a large hysteresis in the electrowetting response to the increasing and decreasing potentials on the PTFE membrane, suggesting an irreversible electrowetting. **(d)** The variation of droplet apparent CA subject to electrowetting cycles. Droplet on the liquid-infused film exhibits nearly perfect electrowetting reversibility with apparent CA switching between ~ 103 ° and ~ 53 ° when the droplet is excited by voltage-on and voltage-off for 10 cycles (blue line). In contrast, the electrowetting on the PTFE membrane is irreversible. Subject to a cyclic voltage-on/voltage-off actuation, the droplet in EWOD attains a switching angle of less than 5 °.

Precautions are taken to prevent the occurrence of unwanted breakdown of the dielectric layer[34,52]. Upon an application of an actuation voltage of 500 V in ambient conditions, the apparent CA of the droplet on the liquid-infused film decreases to ~53 °, corresponding to a switching angle of 50 ° (Fig. 2a). Careful observation of the contact line at the oil-water interface indicates that wetting ridges[44,53,54] are formed owing to the deformation of liquid lubricant rendered by the capillary pressure in electrowetting (Fig. 2b). Meanwhile, by analysis of thermodynamic stability of such a system, it is noticed that the oil would preferentially wet the Teflon matrix, with an extremely thin oil film spreading over and cloaking the water drop (see Supporting Information for details). Moreover, we do not observe the formation of microscopice oil drops entrapped beneath the water drop during electrowetting process[10], presumably due to our unique porous nanostructure, which makes pathway for oil to flow freely inside the porous structure. By contrast, for the case of EWOD where the superhydrophobic PTFE membrane serves as the dielectric layer, the application of an excitation potential of 500 V yields a switching angle of 47 °. Referring to the modified Young-Lippmann equation[5], the dynamic apparent CA ($\theta_E$) of the PTFE membrane corresponding to an applied potential $U$ can be expressed as $\cos\theta_E = \cos\theta_0 + \phi_s \varepsilon U^2 / 2d\gamma_{la}$, where $\theta_0$ is the initial apparent CA, and $\phi_s \varepsilon U^2 / 2d\gamma_{la}$ is the effective electrowetting number. $\phi_s$, $\varepsilon$ and $d$ respectively represent the solid fraction, the dielectric constant, and the thickness of dielectric layer, and $\gamma_{la}$ is the liquid surface tension. Note that although the thickness of the liquid-infused film is 2.5-fold of that of the PTFE membrane, the relatively larger switching angle observed in EWOLF is due to the infiltration of dielectric liquid in the porous matrix, which leads to a significantly enhanced porosity (unity) relative to the porous PTFE membrane. The breakdown voltage of the hybrid liquid-infused film is measured to be 1100 V, over which there is a large breakdown current in the circuit.

The creation of hybrid liquid-liquid interface also provides a pathway for the droplet to freely propagate with little contact line pinning, leading to superior reversibility. Fig. 2c shows the variation of droplet apparent CA under increasing and decreasing potentials. In the experiment, the potential is altered in increments of 50 V every 5 s. As shown in Fig. 2c, at any given applied potential the apparent CA in the receding curve (the potential is decreased) displays nearly perfect fidelity to that in the spreading curve (the potential is increased). As the potential is decreased to 0 V, the droplet turns back to its initial wetting configuration, indicating a complete reversibility. The electrowetting reversibility is further confirmed by measuring the apparent CA subject to electrowetting cycles. As shown in Fig. 2d, we did not observe notable degradation in the switching angle for 10 cycles of voltage-on (500 V) and voltage-off processes. On the contrary, for the superhydrophobic PTFE membrane, there exists a large hysteresis in the electrowetting responses to increasing and decreasing potentials (Fig. 2c), suggesting that the electrowetting on the superhydrophobic PTFE membrane is irreversible. The electrowetting irreversibility is owing to the contact line pinning as a result of liquid penetration into the porous structure in the spreading process (increasing potentials). Dynamic wetting measurement also indicates that the contact angle hysteresis on the PTFE membrane is ~40 °, which is much larger than that on the liquid-infused film (~3 °). Subject to cyclic voltage-on/voltage-off actuation, the observed switching angle in EWOD is less than 5 ° (Fig. 2d).

**Suppressing droplet oscillation in the transient response**. The dynamic response of a droplet in EWOD/EWOLF can be described using a second-order model under a step input. Subject to a sudden external excitation, the droplet typically develops an initial transient response and finally reaches a steady state. For the case of EWOD in our experiment, the transient response of the droplet to a step excitation of 500 V displays two distinctive modes, which are the mobile contact line (MCL) mode and constant contact line (CCL) mode, respectively.

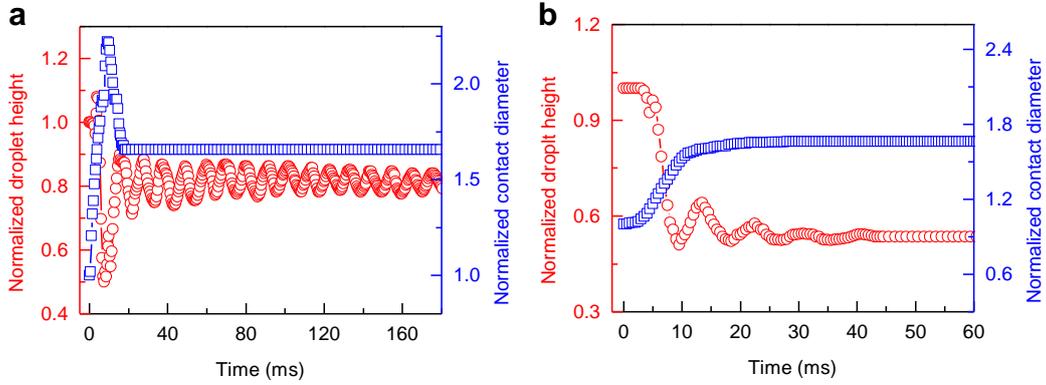

**Figure 3 | Droplet oscillation suppression in the transient response of EWOLF. (a-b)** Transient responses of the activating droplet characterized by the normalized contact diameter ($l/l_0$) and droplet height ($h/h_0$) on the PTFE membrane **(a)** and the liquid infused film **(b)**. During the electrowetting, the transient response of droplet displays two distinctive modes, which are MCL and CCL modes, respectively. In the MCL stage, the maximum contact diameter in EWOD overshoots its final steady-state value, suggesting the under-damped oscillation nature of the activating drop. In the CCL stage, the droplet height attains periodical oscillations. By contrast, for the case of EWOLF, the normalized contact diameter progressively increases without retraction in the MCL stage and droplet oscillation in the droplet height is substantially reduced. Moreover, the response time for the droplet to reach a steady state is ~ 45 ms, representing over 90% reduction compared to that in EWOD. Such a droplet oscillation suppression in the transient response of EWOLF is ascribed to the introduction of viscous liquid lubricant in the porous PTFE matrix, which allows for the formation of additional wetting ridges at the periphery of the droplet as well as viscous liquid-liquid interfaces underneath the droplet.

The MCL mode is characterized by the continuous droplet spreading across the surface (0 ~ 9ms) driven by the electrocapillary pressure, followed by the contact line retraction in the lateral direction. Note that such a lateral motion of droplet, together with the transient vibration of droplet itself, naturally leads to a time-dependent oscillation in the droplet height. Fig. 3a shows the time evolution of contact diameter of droplet ($l$) normalized with respect to its initial contact diameter ($l_0$). It is apparent that the maximum contact diameter of droplet in the MCL stage exceeds its final steady-state value, suggesting an under-damped nature of the vibrating droplet. In the following CCL mode (after 17.5 ms), the contact line is arrested due to CA hysteresis and accordingly, the droplet height exhibits periodical oscillations as reflected by the time evolution of the droplet height ($h$) normalized by the initial droplet height ($h_0$) as shown in Fig. 3a. At the end of the transient response at ~ 500 ms in our observation, the droplet reaches its final steady state. By contrast, the activating droplet in EWOLF yields a distinctively different transient response (Fig. 3b). The droplet contact diameter progressively increases without retraction in the MCL stage and compared to that in EWOD, the oscillation in the droplet height is substantially reduced in the CCL mode. Moreover, the response time for the droplet to reach the steady state is ~ 45 ms, representing over 90% reduction compared to that in EWOD. The droplet oscillation suppression in the transient response in EWOLF is ascribed to the infiltration of viscous liquid lubricant in the porous matrix, which allows for the formation of additional wetting ridges at the periphery of the droplet anchoring the contact lines as well as viscous liquid-liquid interfaces underneath the droplet. Interestingly, the presence of wetting ridges does not sacrifice the electrowetting reversibility enabled by the liquid-liquid interface. This is because, during cyclic activation/deactivation processes, the contact lines anchoring introduced by wetting ridges can be easily released owing to the flexible and soft nature of these wetting ridges. These results suggest that by engineering liquid-liquid interface in EWOLF, we can achieve electrowetting with full reversibility and controlled transient response simultaneously.

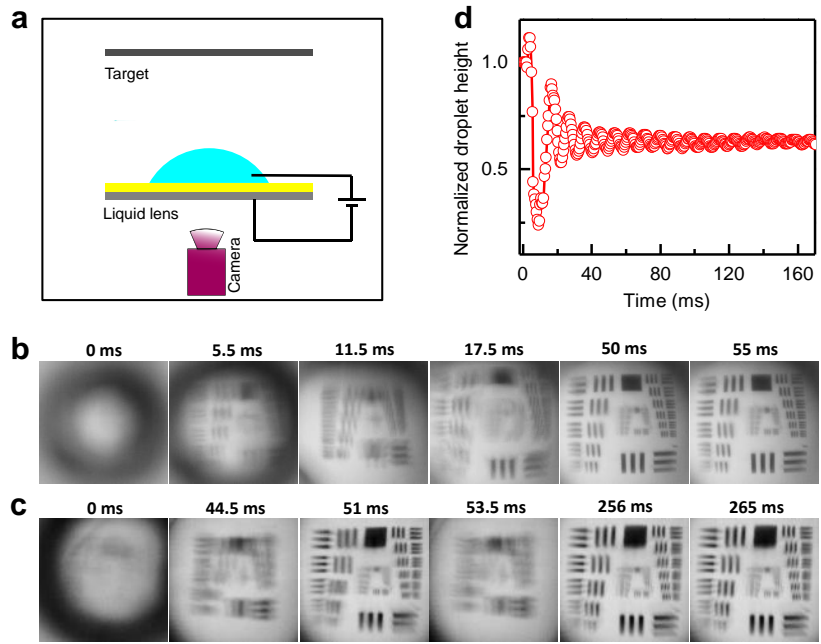

**Figure 4 | Adaptive liquid lens for fast optical focusing.** (**a**) Schematic diagram of EWOLF used for fast imaging experiment. (**b**, **c**) Time lapsed images of a standard resolution target formed through the adaptive liquid lens in EWOLF (**b**) and (**c**) EWOD, respectively. The response time for the EWOLF and EWOD to have sharp and stable images is ~50 and ~250 ms, respectively. When the images captured through the EWOLF have reached a steady state, the images through the EWOD still undergo periodical oscillations, alternating between blurry and sharp states, as indicated by selected snapshots of images within one period of the vibration ranging between ~ 44.5 ms and ~ 53.5 ms. (**d**) The fluctuation in imaging in EWOD is due to the continuous droplet oscillation, as shown by the time variation of the normalized droplet height. The normalized droplet height at 51.0 ms is 0.61905, which is consistent with the steady-state value. However, at 53.5 ms, the droplet height increases to 0.66786, deviating from the equilibrium state for optical focusing and therefore leading to an unfocused image.

**Adaptive liquid lens for fast optical focusing**. We envision that the complete reversibility and droplet oscillation suppression observed in EWOLF can be harnessed to develop novel adaptive liquid lens for fast optical focusing. Fig. 4a shows the schematic drawing of our experimental setup. The liquid lens is placed on the liquid-infused membrane and excited by an external source voltage of 500V. A standard-resolution target (USAF 1951 chart) is positioned parallel to the liquid lens at a distance of ~38 cm, and a high-speed camera (with a frame rate of 2000 fps) is implemented to record the optical image through the liquid lens. At 0 ms, the target is out of focus and the captured image through the liquid lens is blurry (Fig. 4b). When the source voltage is turned on, captured images undergo continuous fluctuation and finally become sharp and stable at ~ 50 ms. With time progression, no further visible change is observed (Fig. 4b). To demonstrate the effect of viscous liquid-liquid interfaces on fast optical focusing, we conducted dynamic imaging based on transparent planar dielectric layers consisting of ~1.8 μm CYTOP coated with ~ 0.95 μm Teflon AF 1600. Careful comparison of captured images in EWOLF and EWOD indicates that when the images captured through the EWOLF have reached a stable state at ~50 ms, images through the EWOD still undergo periodical oscillations, displaying alternative blurry and sharp features. The response time to reach the steady state in the case of EWOD is about four-fold longer than that in EWOLF. Fig.4c shows selected snapshots of images within one period of the vibration ranging between ~ 44.5 ms and 53.5 ms. The image captured at ~ 51 ms is clear and in focus, however, it then switches to a blurry state as a result of continuous droplet oscillation. Such a droplet oscillation is confirmed by the variation of the normalized droplet height over time as shown in Fig. 4d. The normalized droplet height at 51.0 ms is 0.61905, which is consistent with its steady-state value. However, at 53.5 ms, the droplet height increases to 0.66786, deviating from its equilibrium for optical focusing and therefore producing an image that is out of focus. We believe that the simple yet effective approach to achieve fast optical focusing demonstrated here could lead to the development of novel three-dimensional imaging instruments with rapid data collection.

**Discussion**

Previously we have shown the complete reversibility and oscillation suppression in EWOLF as well as its application as liquid lens for fast optical focusing. We ascribed these functionalities to the introduction of liquid-liquid interface, which decreases the CA hysteresis and suppresses severe droplet oscillation encountered in the conventional EWOD. To further elucidate the fundamental mechanism responsible for the controlled oscillation suppression in EWOLF, we first considered the energy dissipation mechanism underlying the EWOD. Without the incorporation of liquid lubricant in the PTFE matrix for the case of EWOD, the energy is mainly dissipated through the contact line friction at the solid-liquid interface and the internal viscous shear caused by the liquid flow inside the droplet. However, owing to the storage of air pockets within porous matrix and low viscosities of water and air, such energy dissipation is minimal and hence the droplet oscillation is considerable. By contrast, replacing air pockets by the viscous liquid lubricant in EWOLF leads to a different energy dissipation mechanism. Compared to the minimal solid-liquid energy dissipation underneath the droplet in EWOD, the viscous liquid-liquid energy dissipation occurring in the MCL and CCL stages is herein increased. Importantly, the infused liquid lubricant is capable of reconfiguring itself, thereby leading to the formation of additional wetting ridges. Thus, compared to the free boundary condition in EWOD, the anchoring effect of wetting ridges naturally contributes to enhanced viscous energy dissipation in the MCL and CCL stages. Taken together, the liquid lubricant serves as a damper, suppressing the occurrence of severe droplet oscillation as observed in EWOD.

To further quantify the energy dissipation associated with the electrowetting process, we calculated the damping ratio. As discussed before, the energy dissipation in the MCL stage in EWOD is primarily governed by the contact line friction and the analysis of the variation of time-dependent normalized contact diameter yields a damping ratio of 0.171. Furthermore, based on the analysis of the transient response of the normalized droplet height, the damping ratio associated with the CCL stage is calculated to be 0.024, which is much smaller than that in the MCL stage. Similarly, the damping ratios in the liquid-infused film (50 μm) in the MCL and CCL modes are calculated to be 1.01 and 0.124, respectively, both of which are larger than those in EWOD.

The damping ratio or the transient response in EWOLF can be tailored by manipulating the viscosity and thickness of liquid lubricant infused in the PTFE membrane. To demonstrate this, we engineered three control surfaces infused with FC-70 and Krytox GPL103 of different thickness (20 μm and 50 μm) and investigated their transient responses. The kinematic viscosity ($\eta$) of FC-70 and Krytox GPL103 is 12 and 80 cSt, respectively. We term the liquid-infused film made of PTFE membrane infused with FC-70 of ~ 50 μm discussed above as 5012, where the first and last two digits refer to the lubricant thickness (unit: μm ) and viscosity (unit: cSt), respectively. Similarly, three control films are termed 2012, 2080 and 5080. Fig. 5a plots the transient responses of the normalized droplet height on three control surfaces. For the electrowetting on surfaces 2080 and 2012, there exists remarkable overshooting in the transient response of the normalized droplet height relative to their individual steady-state values whereas the normalized droplet height on the surface 5080 gradually decreases until reaching its steady state. Note that the damping ratio in the MCL stage is considerably dependent on the thickness and viscosity of liquid lubricant (Fig. 5b). For example, surface 5080 corresponding to the liquid lubricant with the larger viscosity and thickness yields the larger damping ratio of 4.0 (in the MCL stage), which is one order of magnitude larger than that in EWOD. However, because of its almost immobilized contact line, the damping behaviors in EWOLF during the CCL stage become insensitive to the liquid thickness and viscosity, though they are still much larger than that in EWOD. The response time of the droplet to reach the steady state on the surface 5080 is ~ 78 ms, which is much larger than that on surfaces 2012 (~ 65 ms) and 2080 (~ 50 ms), suggesting the importance of the thickness and viscosity of liquid lubricant for the damping control. Note that although reversible electrowetting has been extensively reported through the complete encapsulation of the aqueous droplet in the oil environment[1], the electrowetting system is prone to being over-damped and accordingly leads to a larger response time due to the significantly enlarged liquid-liquid contact area in such configuration. Moreover, once the aqueous droplet is fully encapsulated in the oil phase[55-57], it is challenging to be addressed and poses challenges for subsequent operations.

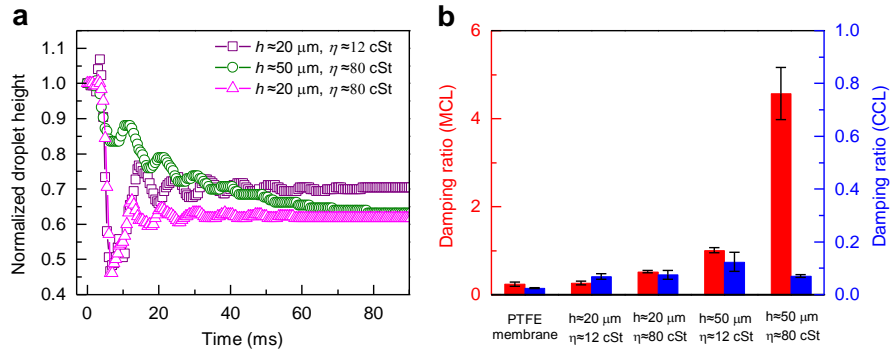

**Figure 5 | Characterization of the transient response and the damping ratio of activating droplets on different liquid-infused films.** (a) Transient responses of normalized droplet height on three control surfaces (2012, 2080, and 5080), corresponding to the PTFE membrane infused with FC-70 of 20 μm, Krytox GPL103 of 20 μm and 50 μm, respectively. For electrowetting on surfaces 2080 and 2012, there exists remarkable overshooting in the transient response of normalized droplet height relative to their individual steady-state values in the MCL stage whereas on the surface 5080 the droplet height gradually decays until reaching its steady state. The response time of droplet to reach the steady state on the surface 5080 is ~ 78 ms, which is much larger than that on surfaces 2012 (~ 65 ms) and 2080 (~ 50 ms). (b) Plots showing the damping ratio of all the surfaces investigated. The damping ratio in the MCL stage exhibits a considerable dependence on the thickness and viscosity of liquid lubricant whereas the damping ratio in the CCL stage in EWOLF becomes almost insensitive to the liquid thickness and viscosity, though they are still much larger than that in EWOD.

In summary, we report on a novel electrtowetting featuring the liquid-infused film as the dielectric layer, which allows for the complete electrowetting reversibility and droplet oscillation suppression without sacrificing each other. Since the actuation principle does not involve moving mechanical parts and current flow, the EWOLF is scalable and of low power consumption. In addition to these merits, the soft liquid-liquid interfaces nature of the EWOLF also makes it practical to develop flexible electrowetting devices for a broad spectrum of applications.

**Methods**
**Sample preparation.** To fabricate the liquid-infused film used for electrowetting, the transparent glass substrate coated with Indium Tin Oxide (ITO) was first cleaned by plasma for ~30 min. After cleaning, the PTFE membrane (thickness ~ 20 μm, Sterlitech Corporation) was gently placed on the top of ITO glass slide. To make the PTFE membrane smooth and be of good contact with the ITO substrate, we wet the membrane using ethanol with the help of capillary wicking effect. Then after the evaporation of ethanol, we added the liquid lubricant with a pre-determined volume into the membrane. In our experiment, the low surface tension 3M Fluorinert FC-70 ($\gamma \approx$ 18 mN/m, $\eta \approx$ 12 cSt) and Dupont Krytox GPL 103 ($\gamma \approx$ 16~20 mN/m, $\eta \approx$ 80 cSt) were chosen as liquid lubricants to spread on PTFE membrane and form a stable film via capillary wicking. The thickness of the liquid-infused film $H$ is determined by the fluid volume $V$ according to the equation $H = (V + m/\rho)/S$, where $\rho$ is the PTFE membrane density, $m$ and $S$ are the weight (measured by analytical balance, AY220, Shimadzu) and the surface area of the membrane, respectively. In our experiment, the thickness of the as-prepared liquid-infused film was tailored to ~ 20 μm and ~ 50 μm, respectively.

**Characterization of wetting properties and electrowetting responses.** We characterized the static wetting properties of the PTFE membrane and liquid-infused film using the sessile droplet method in ambient conditions. The temperature was ~ 20 ℃ and the humidity was ~ 65%. We measured the static apparent CA on both surfaces using the contact angle goniometer (ramé-hart Instrument Corporation). The measured CAs on the PTFE membrane and liquid-infused surface are ~ 152 ° and ~ 103 °, respectively. The contact angle hysteresis of the liquid-infused film was ~ 3 °, which was much smaller than that on the PTFE membrane (~ 40 °).

**Electrowetting experiment.** The electrowetting experiment was conducted in ambient conditions. The potential was applied between the Tungsten tip and ITO electrode. In the experiment, we first examined the variations of apparent CA of droplet on the liquid-infused film under both increasing and decreasing potentials. The potential was altered in increments of 50 V every 5 s. The current was monitored using an ammeter (Keithley 2400). Throughout the experiment, the maximum applied potential was carefully monitored to avoid the breakdown of the dielectric layer. We also examined the variation of apparent CA subject to electrowetting cycles (the applied voltage was turned on and off for 10 times every 10 s). The EWOD responses using the PTFE membrane as the dielectric layer under increasing and decreasing potentials as well as electrowetting cycles were conducted as well.

We also characterized the transient electrowetting responses to a step actuation voltage of 500 V in EWOD and EWOLF. We used the high speed camera (FASTCAM SA4, Photron) with the frame rate of 2000 fps to capture the droplet spreading and receding dynamics. The time variations of the normalized droplet height and contact diameter were measured using the image process software, ImageJ.

**Acknowledgements**
This work was supported by the Natural Science Foundation of China General Program (51475401), RGC General Research Fund (9041809), and Natural Science Foundation of China Major Program (91215302), the Industrial Donation Grant (9220041), the National Basic Research Program of China (2012CB933302). We thank Prof. Minhang Bao and Dr. Lei Xu for many helpful discussions.

**Author contributions**
Z.W. conceived the research. Z.W. and C.H. designed the research, C.H., Y.L., and X.C. carried out the experiments, Y.H., Q.L. and K.Y.L. contributed to the data analysis. All the authors contributed to writing and revising the manuscript, and agreed on its final contents.

**Competing financial interests**

The authors declare no competing financial interests.

Chonglei Hao[1], Yahua Liu[1], Xuemei Chen[1], Yuncheng He[2], Qiusheng Li[2], K.Y. Li[1], Zuankai Wang[1, *]

[1]Department of Mechanical and Biomedical Engineering, City University of Hong Kong, Hong Kong, China
[2]Department of Civil and Architectural Engineering, City University of Hong Kong, Hong Kong, China

*Corresponding author. E-mail: zuanwang@cityu.edu.hk

**Drop stable configuration on lubricant infused film.**

To determine the stable configuration for the water droplet placed on lubricant infused film, we first measured the contact angle of oil drop (FC-70 and Krytox 103) on the smooth Teflon coating surfaces under air ($\theta_{os(a)}$) and the water environment ($\theta_{os(w)}$), where subscripts *o*, *s*, *a*, and *w* represent for oil, solid, air and water, respectively[1,2]. Fig. S1 shows the time evolution of FC-70 drop spreading in the air **(a)** and water **(b)** environments. It is obvious that the apparent contact angle in the air ($\theta_{os(a)}$) and water ($\theta_{os(w)}$) environment is close to 0 °, suggesting that the oil can preferentially wet the Teflon matrix and the deposited droplet is floating on the oil-infused film. We further calculated the spreading parameter $S = \gamma_{wa} - (\gamma_{wo} + \gamma_{oa})$, where $\gamma_{wa}$, $\gamma_{wo}$, and $\gamma_{oa}$ are the surface tensions of water-air, water-oil, and oil-air interfaces, respectively. As listed in Table S1, the spreading parameters of two oils on water are both larger than 0, revealing that a thin oil film will cloak the water drop at zero voltage, as shown in Fig. S2.

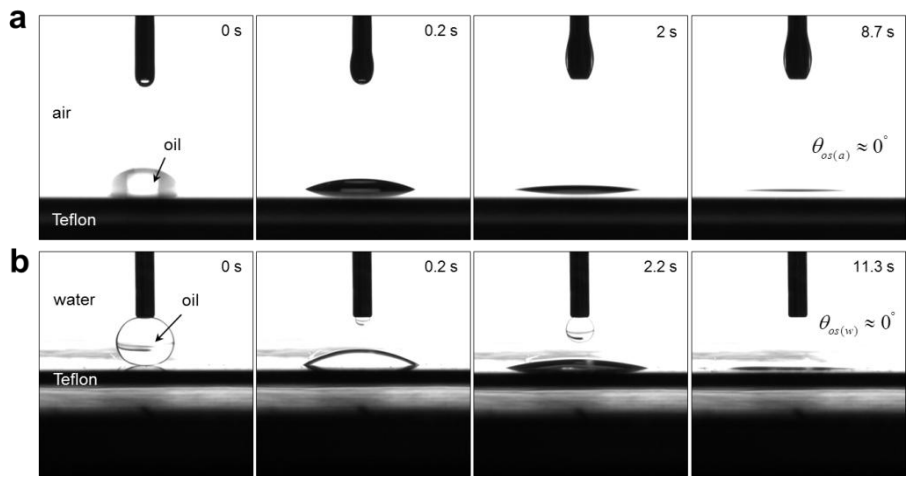

**Figure S1:** Time evolution of FC-70 drop morphology placed on smooth Teflon coating surface in the air **(a)** and water **(b)** environment, respectively. Both of the final contact angles, $\theta_{os(a)}$ and $\theta_{os(w)}$, were nearly 0 °.

**Table S1**. Physical properties of oil and the calculated spreading parameter.

| oil | Surface tension, | Surface tension | Spreading parameter, |
| --- | --- | --- | --- |

|          | $\gamma_{oa}$, mN/m | $\gamma_{wo}$, mN/m | $S$, mN/m |
|----------|---------------------|---------------------|-----------|
| FC-70    | 16.2                | 51.4±1              | 4.4±1     |
| Krytox103| 18.4                | 47±1                | 6.6±1     |

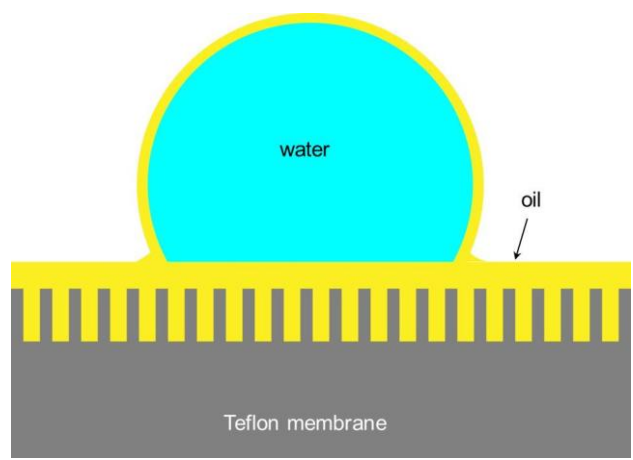

**Figure S2:** Schematic diagram of a water drop placed on the lubricant infused film demonstrates the floating state of water drop.

References

1. Smith, J. D., et al. Droplet mobility on lubricant-impregnated surfaces. *Soft Matter* **9**, 1772-1780 (2013).
2. Wong, T. S., et al. Bioinspired self-repairing slippery surfaces with pressure-stable omniphobicity. *Nature* **477**, 443-447 (2011).